\documentclass[aps,prl,groupedaddress,fleqn,twocolumn]{revtex4}
\pdfoutput=1
\usepackage{graphicx}
\usepackage{xcolor}
\usepackage{amsmath}
\usepackage{gensymb}
\usepackage{float}
\usepackage{physics}
\begin{document}
\title{Pronounced structural crossover in water at supercritical pressures}
\author{C. Cockrell$^1$, O. A. Dicks$^1$, V. V. Brazhkin$^2$, and K. Trachenko$^1$}
\address{$^1$ School of Physics and Astronomy, Queen Mary University of London, Mile End Road, London, E1 4NS, UK
\\ $^2$ Institute for High Pressure Physics, RAS, 108840, Moscow, Russia}


\begin{abstract}
There have been ample studies of the many phases of H2O in both its solid and low pressure liquid states, and the transitions between them. Using molecular dynamics simulations we address the hitherto unexplored deeply supercritical pressures, where no qualitative transitions are thought to take place and where all properties are expected to vary smoothly. On the basis of these simulations we predict that water  at supercritical pressures undergoes a structural crossover across the Frenkel line at pressures as high as 45 times the critical pressure. This provides a new insight into the water phase diagram and establishes a link between the structural and dynamical properties of supercritical water. Specifically, the crossover is demonstrated by a sharp and pronounced at low pressures, and smooth at high pressures, signified by changes in the pair distribution functions and local coordination which coincide with the dynamical transition (the loss of all oscillatory molecular motion) at the Frenkel line on the phase diagram.


\end{abstract}

\maketitle

\section{Introduction}

H$_2$O is arguably the most studied compound. Its properties in crystalline, amorphous, liquid, and supercooled states are well documented, yet not well understood due to a variety of anomalies that continue to inspire enquiry\cite{stanley0,gallo0}. Little is known about the properties of supercritical water despite its increasing deployment in important industrial and environmental applications \cite{deben,akiya,savage,huelsman}. Here, we extend pressure far beyond the critical pressure using molecular dynamics (MD) simulations and find that water undergoes a pronounced structural crossover at pressures as high as 45 times the critical pressure. These pressures are far away from the melting line and critical point, corresponding to a part of the phase diagram where, according to traditional view, properties were expected to vary smoothly with no qualitative changes \cite{deben}. The structural crossover at low pressures is defined by the transition from a tetrahedral-like to a more closely-packed molecular arrangement with an accompanying shift in its structural evolution with temperature. The crossover at higher pressures is more subtle, but can be seen in the evolution of the pair distribution functions with temperature. Importantly, both of these changes exactly coincide within a small range of the dynamical crossover across the Frenkel line (FL) proposed previously \cite{phystoday,prl,ropp} and demonstrate that the structural crossover is coupled to the disappearance of transverse modes in supercritical pressure water. Our results give new insight into water's phase diagram, serve as a guide for future high-pressure experiments, and have practical applications as dissolution and extraction properties are optimised at the FL \cite{pre1}.

\begin{figure}
\begin{center}
\includegraphics[width=1.0\linewidth]{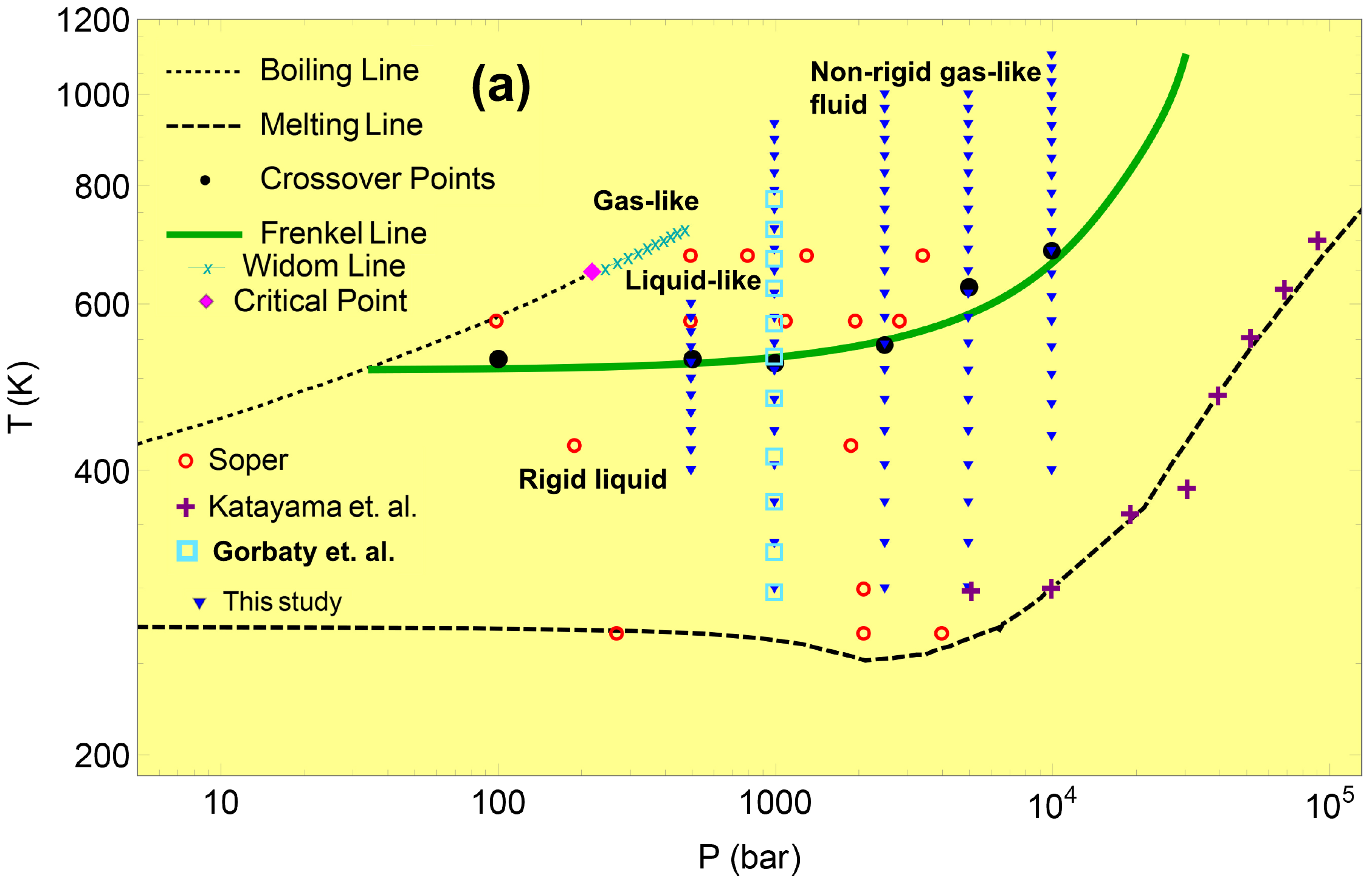}
\includegraphics[width=1.0\linewidth]{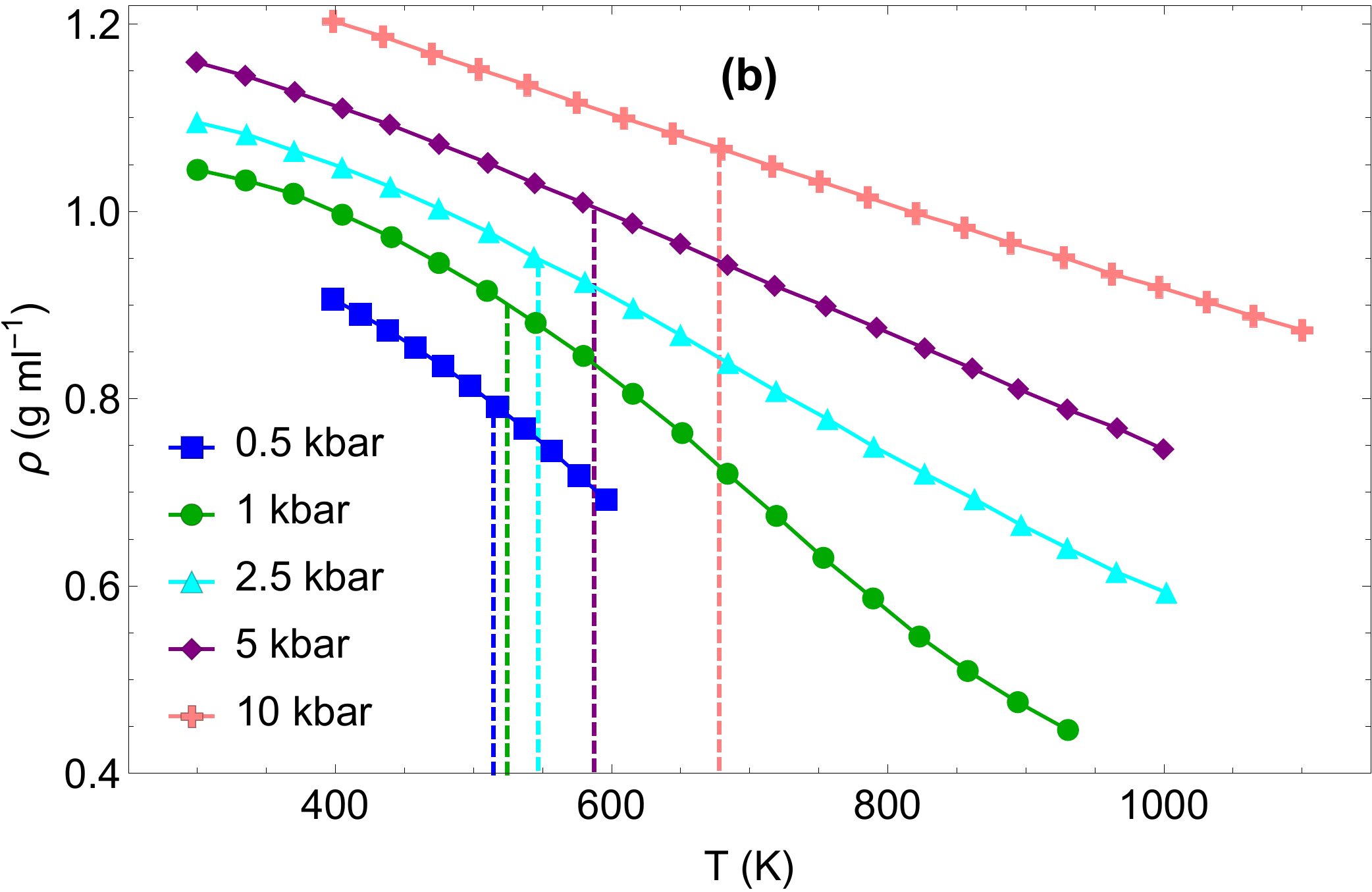}
\end{center}
\caption{(Colour online) (a): Pressure-temperature ($P$,$T$) phase diagram of H$_2$O showing the Frenkel line (reproduced from \cite{pre1}), together with earlier experimental ($P$,$T$) points \cite{katayama,soper,Gorbatyisobar} and currently used state points. We show the Widom line using the data of Ref. \cite{gallo}. The Widom line separates ``gas-like" and ``liquid-like" regions of the near-critical phase diagram, whereas the Frenkel line separates dynamically distinct rigid and non-rigid fluids. (b): Density versus temperature plots for the simulated samples in this study. The Frenkel line passes through regions of high density beneath the density fluctuations at the Widom line.}
\label{diagram}
\end{figure}

We note that high pressure and temperature experiments in water are challenging and scarce as a result. The structure of high-pressure water was studied along the melting curve \cite{katayama}, however very few studies explored conditions both above the melting curve and close to the FL \cite{soper,Gorbatyisobar}. These experimental challenges resulted in a widely-spaced distribution on the temperature-pressure phase diagram as shown in Fig. \ref{diagram}. Coupled with no guide from theory, this precluded the identification of the FL crossover in water.

Traditionally the deep supercritical state was thought to undergo only smooth changes in response to pressure and temperature without any qualitative changes \cite{deben}. Recent discussions challenged this understanding. Close to the critical point, a demarcation of the supercritical state was proposed on the basis of the Widom line (WL). This is the line of critical anomalies persisting beyond the critical point, defined as the line of maximum of properties such as heat capacity or correlation length \cite{stanley}. Another demarcation of the supercritical state, the Frenkel line, has been introduced, instead based on qualitative changes of particle dynamics \cite{phystoday,prl,ropp}. Below the line, dynamics combine solid-like oscillations around quasi-equilibrium positions and diffusive jumps between different positions. Above the line, particle dynamics lose the oscillatory component and become purely diffusive \cite{phystoday,prl,ropp}. This gives a practical criterion to calculate the FL based on the disappearance of the minima of the velocity autocorrelation function (VAF). The FL corresponds to the loss of solid-like transverse quasi-harmonic modes from the system spectrum, corresponding to the specific heat $c_v$ equal to $2k_{\rm B}$ in the harmonic case for simple systems. This represents another, thermodynamic, criterion of the FL, which gives the same line as the VAF criterion \cite{prl}. Differently from the WL, the FL is (a) unrelated to the critical point and exists in systems without it, (b) extends to arbitrarily high pressure and temperature if chemical bonding is unaltered (the WL disappears above the critical point fairly quickly as is seen in Fig. 1) and (c) independent on the path taken on the phase diagram \cite{phystoday,prl,ropp}.

Since structure and dynamics are related, the dynamical crossover should result in a change in the evolution of structure with temperature \cite{ling}. We find that the structural crossover at the FL in water at deeply supercritical pressures is sharp and pronounced.

Interesting and anomalous effects in water are related to its tetrahedral structure and transitions to higher-coordinated states, as discussed below. At GPa pressures, water becomes high-coordinated and its tetrahedral network is lost \cite{katayama}. Below 0.5 kbar, water forms a tetrahedral network at low temperature, but its properties are affected by persisting near-critical anomalies, obscuring structure-related effects. This gives an interesting unexplored window around 1--5 kbar where water is tetrahedral at low temperature and is unaffected by the vicinity of the critical point.

The FL for H$_2$O was previously calculated using the VAF criterion \cite{pre1} using the TIP4P/2005 potential that we employ here. The FL calculated from this method is reproduced in Fig. \ref{diagram}(a). This gives the following state points at the FL in 0.5--10 kbar pressure range: (0.5 kbar, 515 K), (1 kbar, 525 K), (2.5 kbar, 550 K), (5 kbar, 580 K), and (10 kbar, 680 K). The FL extends to arbitrarily high pressure and temperature above the critical point, but at low temperature it terminates at the boiling line at around 0.8$T_c$, where $T_c$ is the critical temperature \cite{prl} (note that the system lacks cohesive liquid-like states at temperatures above approximately 0.8$T_c$ \cite{stishov}, hence crossing the boiling line at those conditions can be viewed as a gas-gas transition \cite{prl}.) Water's critical point is $P_c=0.22$ kbar, $T_c=647$ K, hence the first four state points are above $P_c$ and below $T_c$, whereas the last one is above both $P_c$ and $T_c$. These state points therefore correspond to temperatures much higher than the melting temperature and pressures far in excess of $P_c$.

Although the sharp crossover we detect is below $T_c$, it is nevertheless far above $P_c$. From the physical perspective, this is notable because the crossover operates far from the boiling line and the critical point, corresponding to the part of the phase diagram where (a) all properties were assumed to vary smoothly with no qualitative changes \cite{deben} and (b) near-critical anomalies are non-existent \cite{uspehi}. Fig. \ref{diagram}(b) plots the density of our simulated samples and shows that, where they exist, the near-critical anomalies lie above the FL and that the FL therefore lies in the ``liquid-like" phase below the Widom line and not the expanding ``gas-like" phase above it.  In this regard, we note that the common definition of the supercritical state (rectangle defined by $P>P_c$ and $T>T_c$) is loose, not least because an isotherm drawn on the ($P$,$T$) diagram above the critical point crosses the melting line (see Fig. 1), implying that the supercritical state can be found in the solid phase (see Ref. \cite{uspehi} for details).

\section{Methods}

\begin{figure}
\begin{center}
{\scalebox{0.6}{\includegraphics{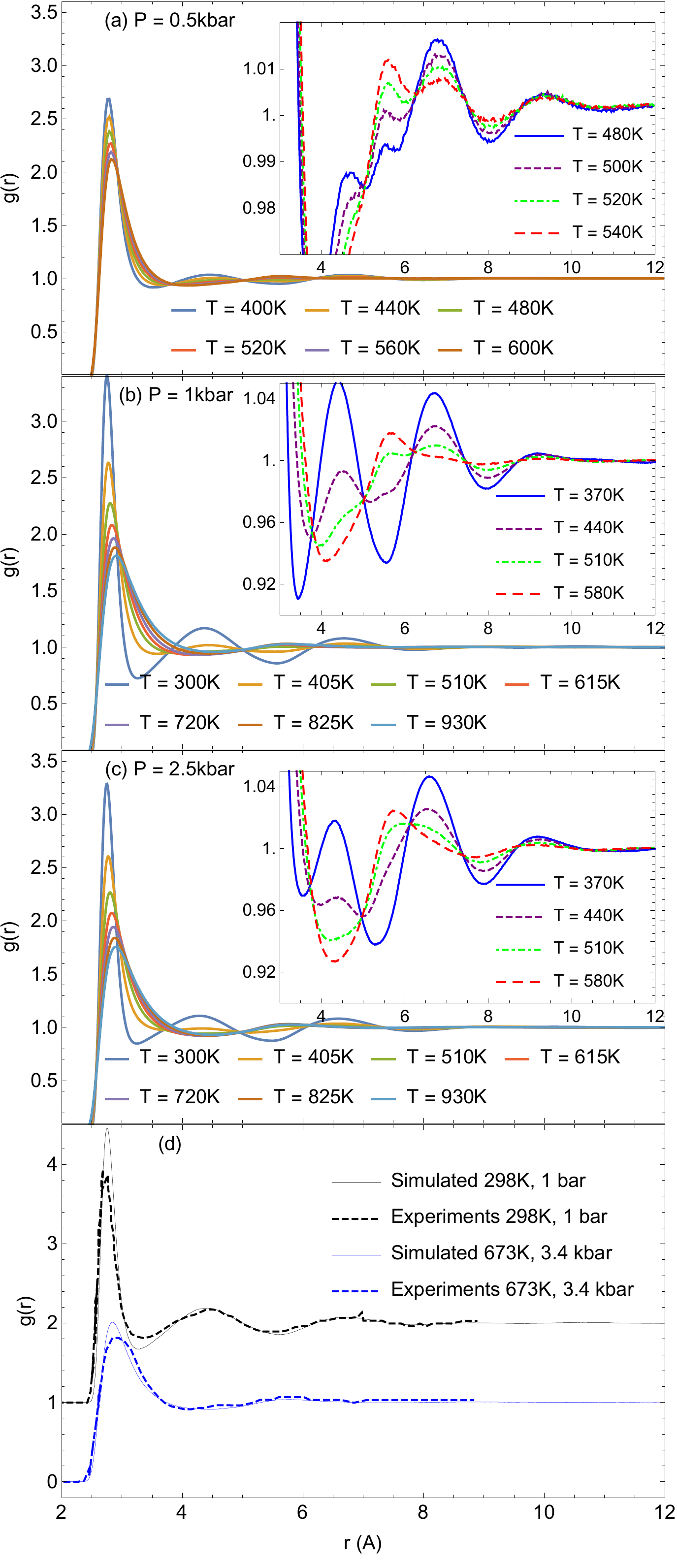}}}
\end{center}
\caption{(Colour online) (a)-(c): O-O PDFs of simulated water at different pressures and temperatures. (d) Simulated and experimental \cite{soper} PDFs at ambient and supercritical conditions, offset by 1 for convenience.}
\label{pdfs}
\end{figure}

\begin{figure}
\begin{center}
{\scalebox{0.55}{\includegraphics{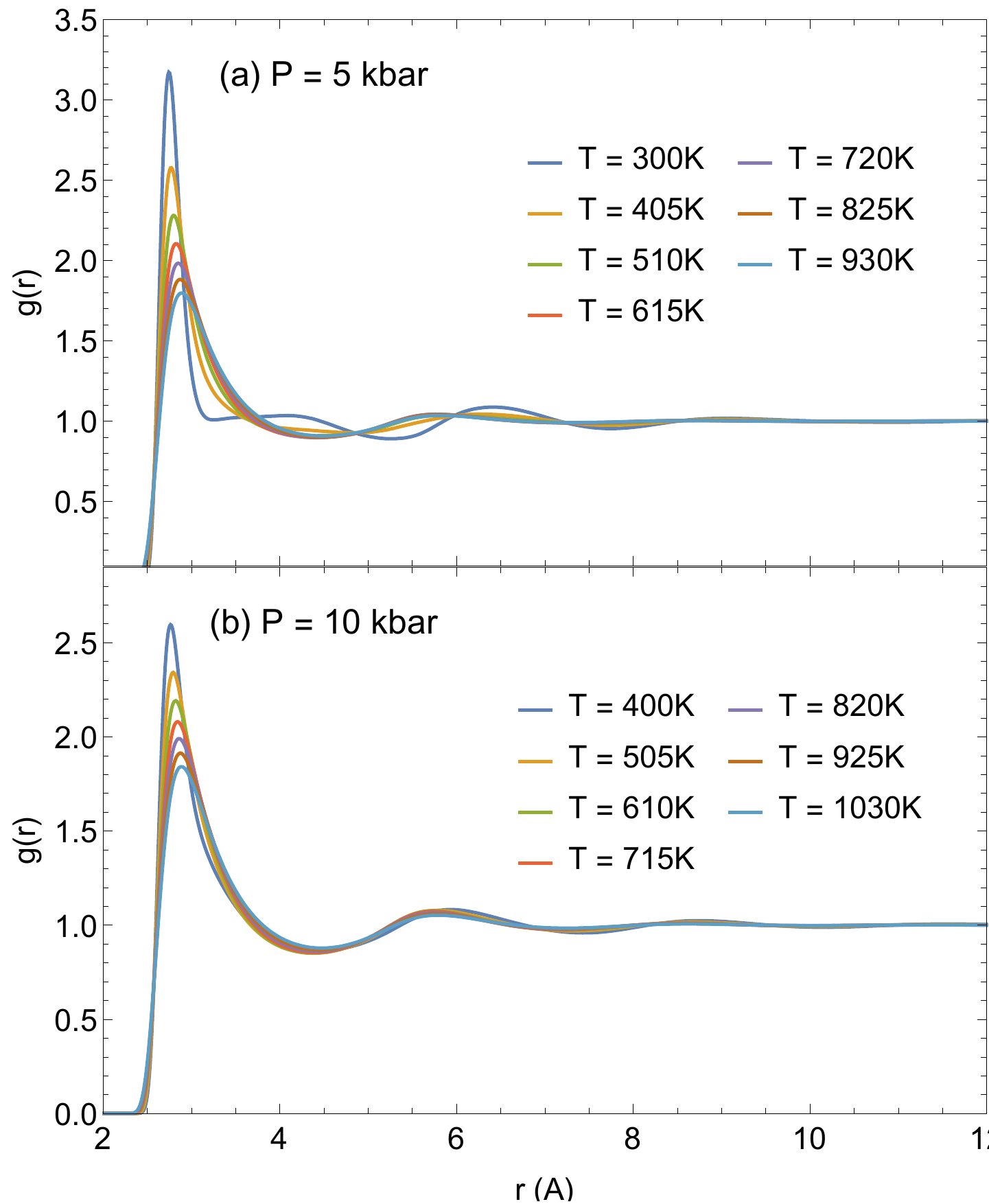}}}
\end{center}
\caption{(Colour online) (a)-(b): O-O PDFs of simulated water at higher pressures.}
\label{pdfshigh}
\end{figure}

We perform MD simulations using the DL\_POLY package \cite{dlpoly} and the TIP4P/2005 potential for water, which is optimised for high pressure and temperature conditions \cite{potential}. A careful analysis \cite{vega,vega1} assigned this potential the highest score in terms of the extent to which the results agree with different experimental properties, including the equation of state, high pressure and temperature behaviour, and structure. This potential was also used in a high pressure and temperature study of the WL in supercritical water \cite{gallo}.

We equilibrated systems of 32768 water molecules in the constant temperature and pressure ensemble at the chosen pressures for 30 ps. The data were collected from subsequent production runs in the constant energy and volume ensemble for 170 ps. We simulated several temperature points at each pressure in a range enveloping the FL (see Fig. \ref{diagram}). Electrostatic interactions were handled by the smooth particle mesh Ewald method.

We also performed simulations of water using the SPC/E potential \cite{spce}, which corroborated our findings and demonstrated that the observed behaviour is not an artefact of the TIP4P/2005 potential.

\section{Results and Discussion}

\begin{figure}[H]
\begin{center}
{\scalebox{0.5}{\includegraphics{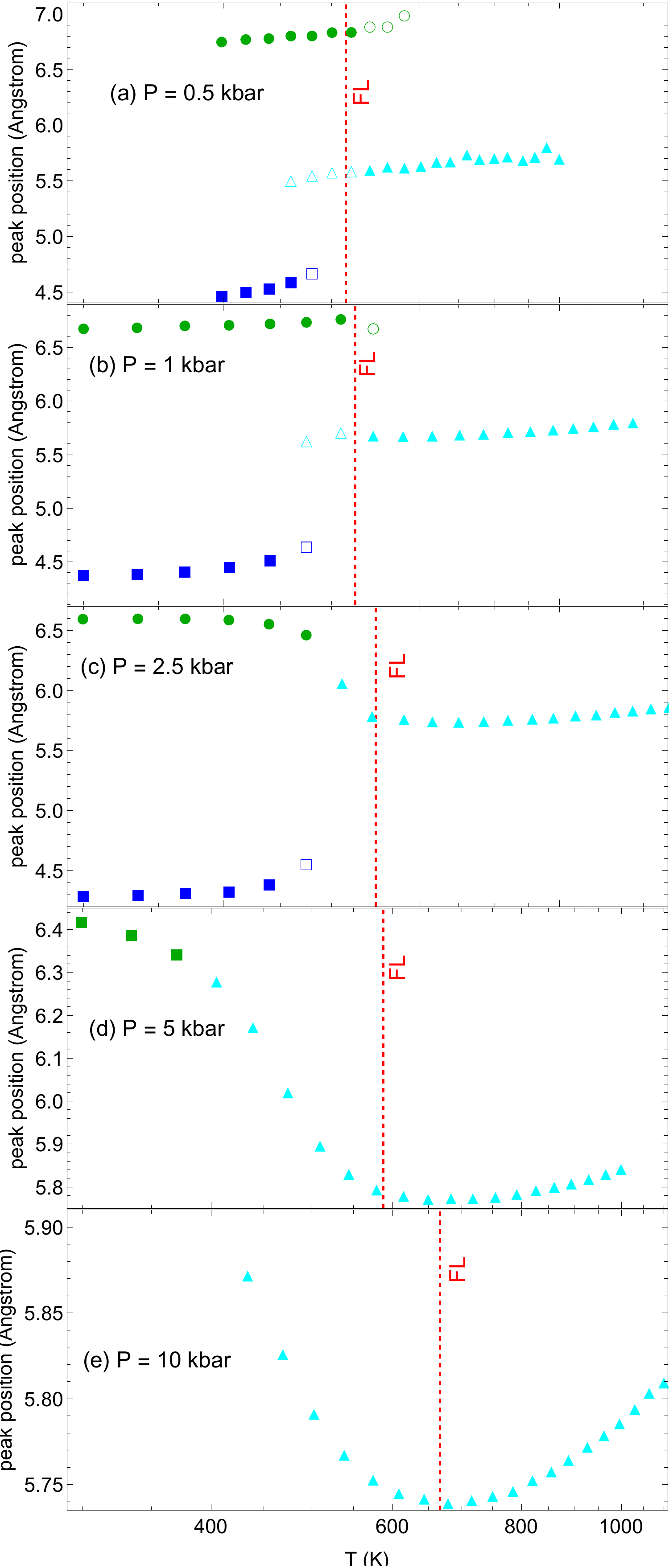}}}
\end{center}
\caption{O-O PDF peak positions: Squares - second peaks; circles - third peak; triangles - new second peaks. Open triangles imply that the new peak is less prominent than the old peaks, and vice versa for open squares and circles.} The dashed vertical lines correspond to temperatures at the Frenkel line.
\label{peakpos}
\end{figure}

We show the calculated pair distribution functions (PDFs) in Figs. \ref{pdfs} and \ref{pdfshigh} (the interatomic potential treats H$_2$O molecules as rigid units, we show O-O correlations). We observe a pronounced structural crossover: at 0.5, and 1.0 kbar we see the disappearance of the second and third peaks as the temperature approaches the FL, (see Fig. \ref{pdfs}). Concomitantly, a new second peak emerges at a new radial position as these peaks diminish. The temperature at which the new peak becomes more prominent than the old peaks coincides very closely with the temperature at the FL, $T_{\rm F}$.  The sharpness of the crossover is most easily observed in the peak positions at $T_{\rm F}$ in Fig. \ref{peakpos}: the second and third peaks become less prominent than the new peak at the FL and are absorbed by neighbouring peaks beyond the FL. This crossover is pronounced in the sense that the second and third peaks do not continuously shift to new positions, but rather they give way entirely to the new second peak. At higher pressures, the crossover is less pronounced for reasons discussed below. At 2.5 kbar, the new peak develops as a shoulder to the third peak, causing the third peak radial position to sharply drop to the new peak position at the FL. At 5 and 10 kbar the second peak has disappeared well below the FL, and the third peak radial position drops more smoothly, reaching a minimum near the FL before increasing again as temperature increases.

\begin{figure}
\begin{center}
{\scalebox{0.5}{\includegraphics{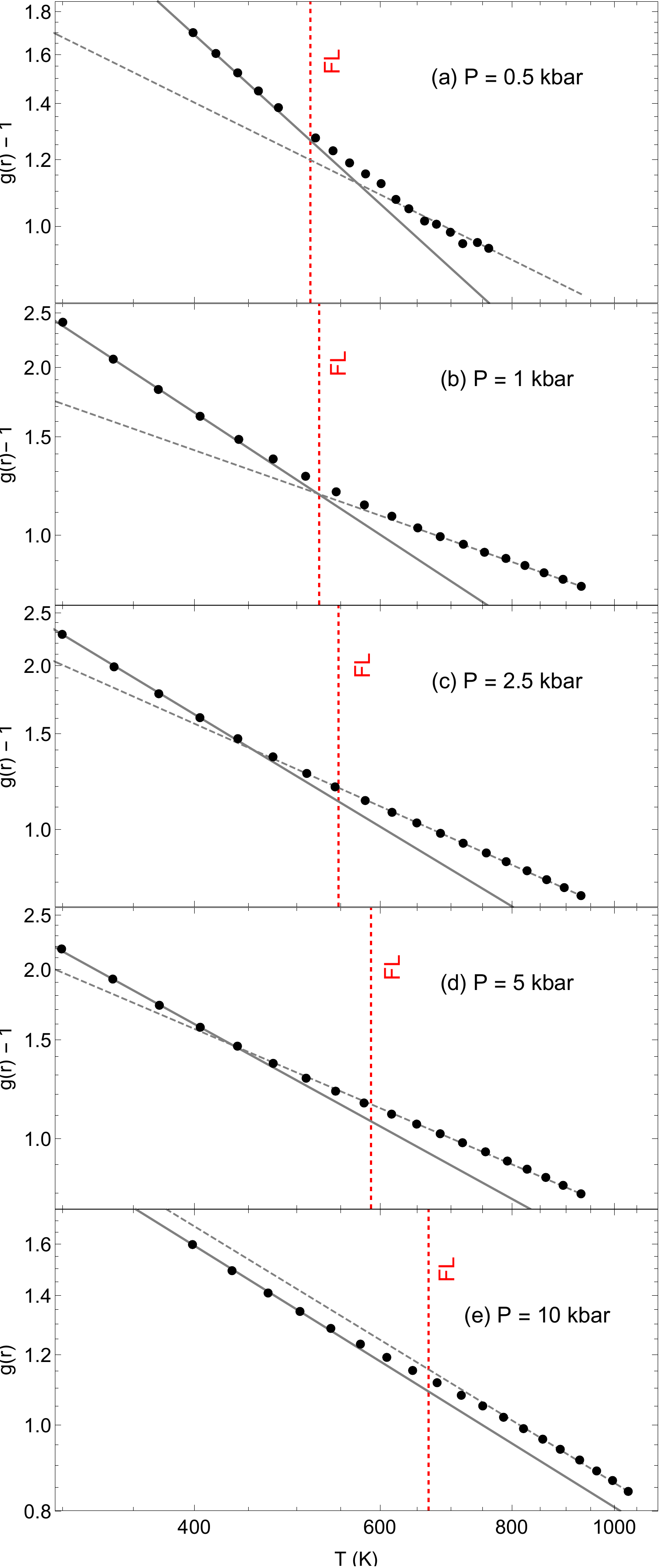}}}
\end{center}
\caption{Log-log plot of the PDF first peak heights, showing the crossover of evolution as temperature approaches the FL. The dashed lines correspond to the temperatures at the FL.}
\label{peakheights}
\end{figure}

We plot the first peak height $h = g(r_{\rm{max}}) - 1$ in Fig. \ref{peakheights}, where $r_{\rm{max}}$ is the peak radial position. We note that the PDF peak heights $h$ of a solid are predicted \cite{tld, frenkel} to have a power-law relationship with temperature: $\log h \propto - \log T$. The same relation should apply to liquids below the FL where the solid-like oscillatory component of molecular motion is present \cite{ling}. For small displacements the energy is roughly quadratic and the displacement distribution will be Gaussian. The height of a Gaussian distribution follows a power-law relationship with its variance, and thus with temperature. The peak heights in Fig. \ref{peakheights} clearly show the crossover at the FL, with the observed crossover temperatures differing from the predicted ones by about 5-15\%. This is in agreement with the width of the FL crossover seen experimentally and modelling on the basis of structural and thermodynamic properties \cite{ne,cv,co2}.

As discussed, experimental PDFs at conditions close to the FL are scarce. We have selected two state points in the experimental work \cite{soper} for direct comparison, one close to ambient conditions and the other at high pressure and temperature far above the melting line and also above the FL. We show the experimental PDFs, together with the PDFs simulated at the same state points, in Fig. \ref{pdfs}d. We observe a good agreement between experimental and simulated O-O PDFs at ambient conditions (a slight overestimation of the first peak height in simulations is a known feature of the potential \cite{vega1}). At high pressure and temperature, we observe a similar behaviour in the experimental data to that seen in Fig. \ref{pdfs}a-b: the second and third peaks have disappeared, and the broad new second peak has emerged at around 6 \AA. Gorbaty \textit{et. al.} performed extensive X-ray scattering experiments on water at the 1 kbar isobar and 293K isotherm. These systematic measurements also exhibit the appearance of the new second peak at around 6 \AA, at temperatures very close to the predicted FL \cite{Gorbatyisobar}. Our results also reproduce the diminishment of the second peak and the appearance of a pronounced shoulder on the first peak along our isotherm which the authors discussed \cite{Gorbatyisotherm}. This gives us confidence in the model's good structural performance in the range of pressure and temperature where we predict the transition.

We attribute the observed crossovers of PDF features to the dynamical crossover at the FL, coupled with a water-specific structural transformation. As discussed, the FL corresponds to the dynamical crossover of molecular motion from combined diffusion and oscillation to pure diffusion. The oscillatory component implies that average molecular positions do not change during time $\tau$, the liquid relaxation time \cite{frenkel}. On the other hand, purely diffusive motion implies continuous molecular rearrangements. As a result, structural correlations are also expected to undergo a crossover at the FL. In Fig. \ref{diagram} the phase points of the crossover closely trace the FL. At low pressures when the crossover is sharp, the crossover point is defined as the point at which the new peak is more prominent than the old peaks. At higher pressures, the crossover point is defined as the temperature at which the new second peak reaches its minimum radial distance, above which the radial distance starts to increase with temperature. This coincides with the crossover in peak heights in Fig. \ref{peakheights}.

In water, this results in the pronounced crossover of the second and third peaks of PDFs for the following reason. Water is known to undergo a structural transformation from a tetrahedral-like structure, governed by hydrogen bonding, at low temperature to a more closely-packed structure at high temperatures and pressures \cite{ikeda,water-tr,marti}. The second peak in the low-temperature PDFs (when structure is tetrahedral-like), corresponding to next-nearest neighbours, disappears during this transformation. In the higher-coordinated structure at high temperature, the second peak corresponds to a new distance which is between the second and the third peaks in the low-temperature structure (see Fig. \ref{peakpos}). This behaviour was seen in subcritical water in quantum-mechanical calculations, and experiments \cite{katayama,soper}. It was also seen in sub- and supercritical water using MC simulations when crossing the FL over isochores \cite{Kalinichev}, demonstrating that this crossover is not dependent on the path taken on the phase diagram.

Based on these observations we propose that the FL {\it facilitates} water's structural crossover between the tetrahedral-like and more closely-packed structures from the near critical state to deep supercritical pressure. As the oscillatory component of molecular motion is lost in the tetrahedral structure, water molecules acquire purely diffusive motion and hence flexibility to arrange into a denser structure in response to high pressure. At low pressures, therefore, water transitions from a tetrahedral rigid liquid to a close-packed rigid liquid below the FL, and from a close-packed rigid liquid to a ``gas-like" non-rigid liquid above the FL.

We emphasize that although the transformation between tetrahedral-like and close-packed structure in water has been discussed before, the novelty here is that this transition is \textit{coupled to the dynamical transition at the FL} and operates at deep supercritical pressure where such transitions were precluded according to the existing picture of supercritical matter as featureless, homogeneous, and lacking any transitions \cite{deben}.

\begin{figure}
\begin{center}
{\scalebox{0.43}{\includegraphics{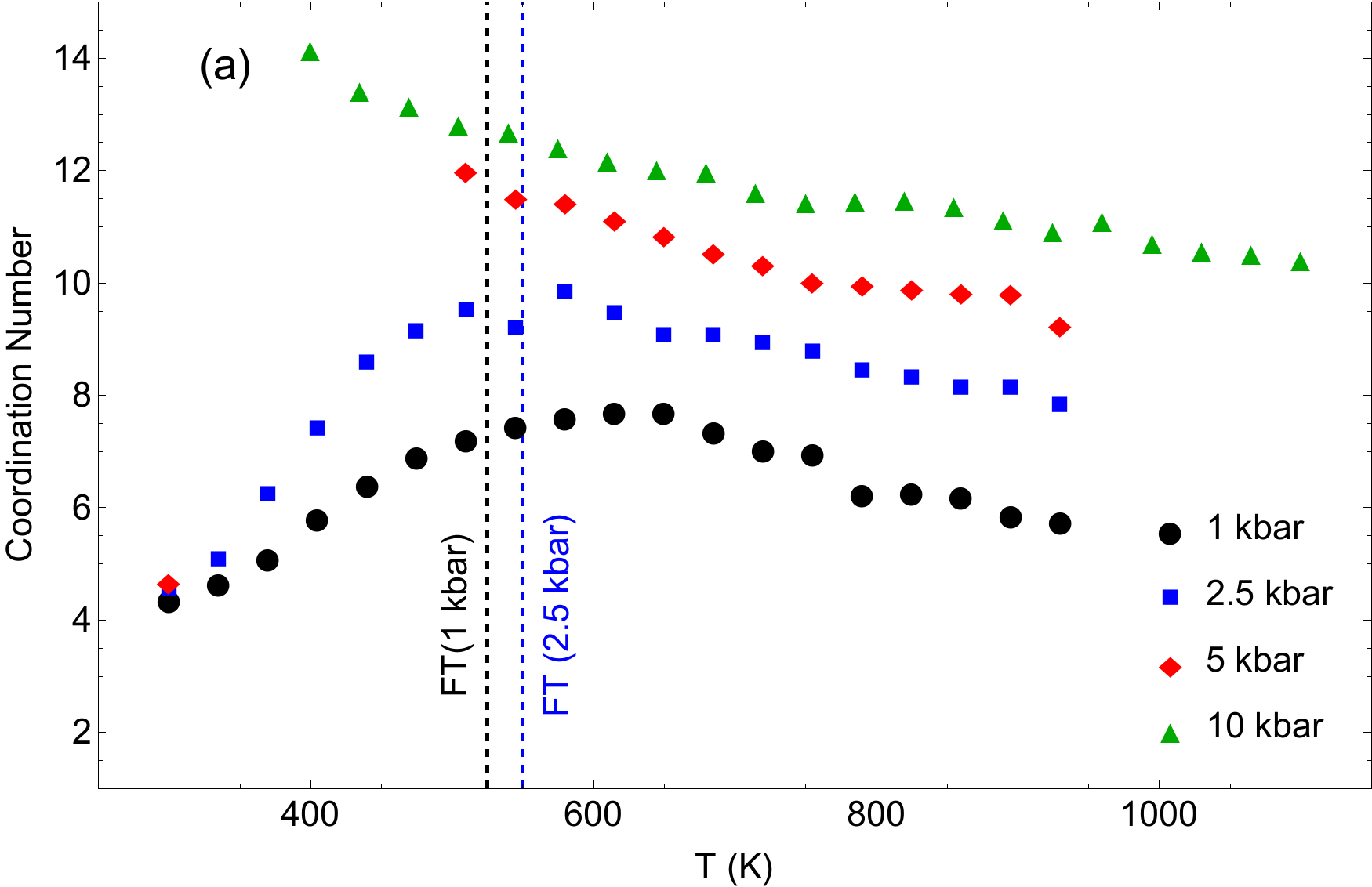}}}
{\scalebox{0.41}{\includegraphics{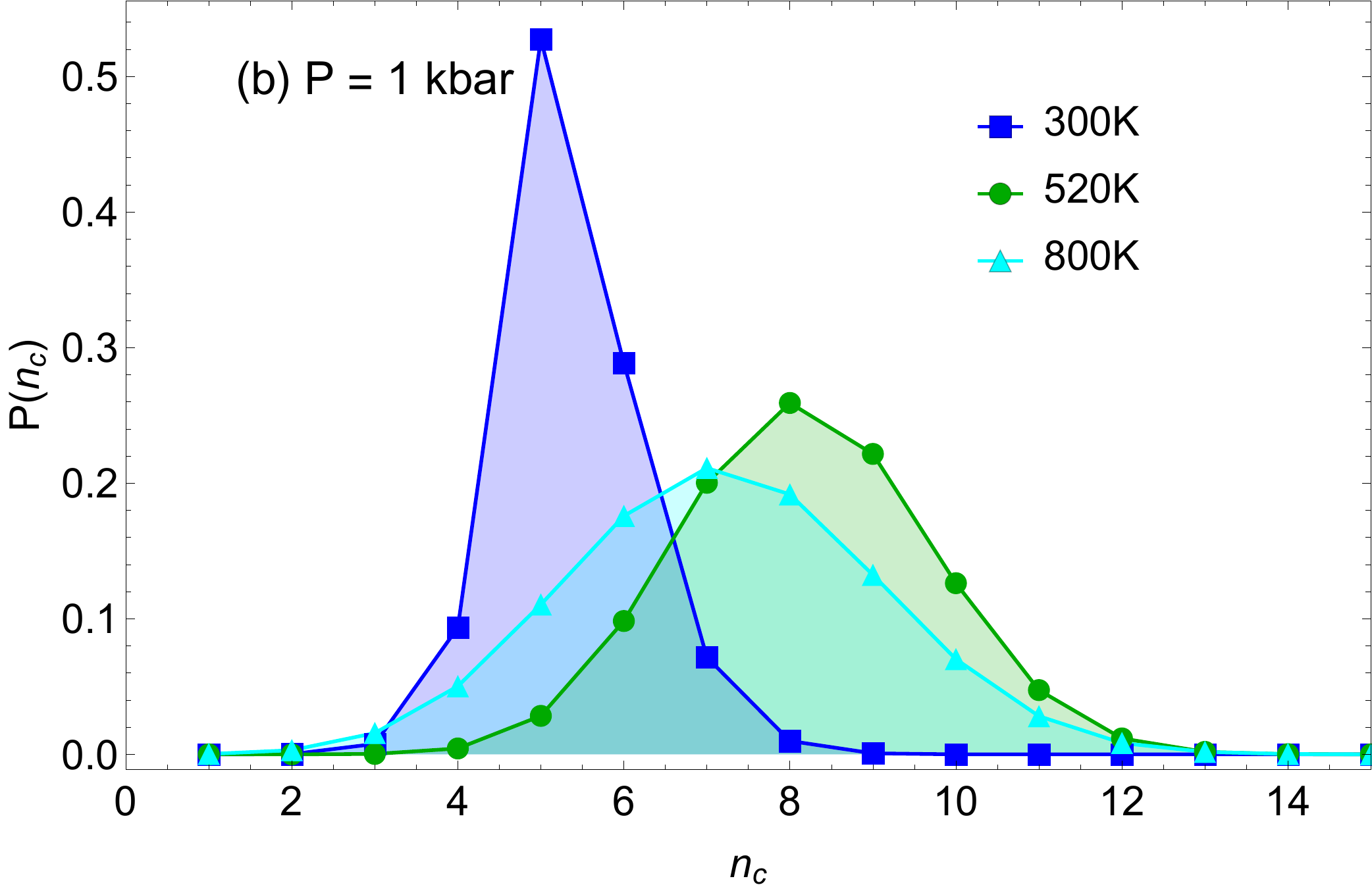}}}
{\scalebox{0.41}{\includegraphics{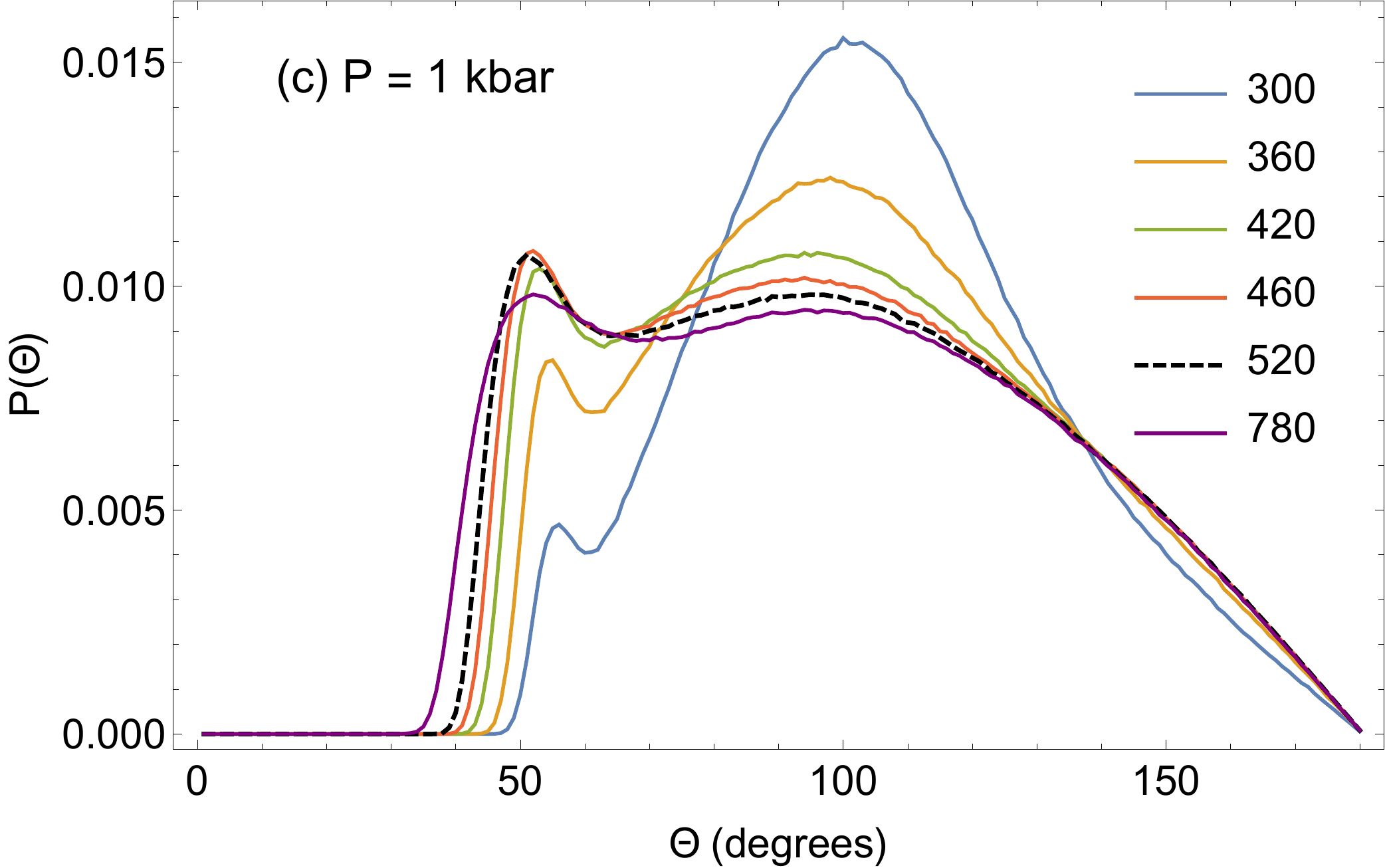}}}
\end{center}
\caption{(a): Average coordination number $n_c$ of water molecules at supercritical pressures as temperatures cross the Frenkel line. $n_c$ is not shown at low temperature at 5 kbar because the minimum between the first and second peaks in Fig. \ref{pdfs}c disappears at those temperatures, causing an ill-defined cut-off. (b): Normalised histogram of molecular coordination calculated from structural snapshots. (c): Intermolecular angular distribution functions at 1 kbar. The dashed curve shows the distribution at the temperature corresponding to the Frenkel line. }
\label{coord}
\end{figure}

\begin{figure}
\begin{center}
{\scalebox{0.11}{\includegraphics{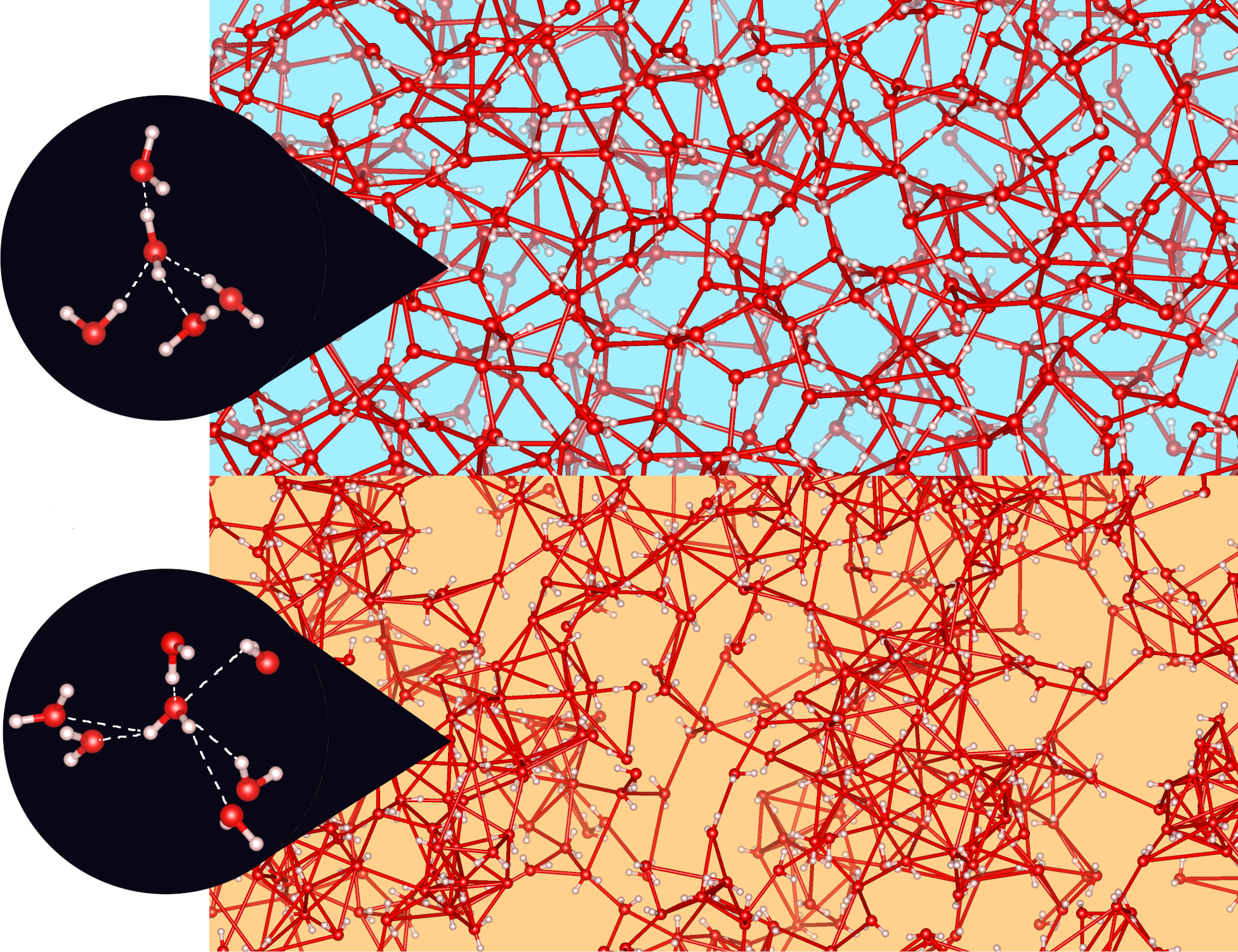}}}
\end{center}
\caption{Snapshots of the structure of simulated water at ($P=1$ kbar, $T=300$ K) (top) and ($P=1$ kbar, $T=930$ K) (bottom) showing 4- and 6-fold coordinated water molecules.}
\label{cartoon}
\end{figure}

We further support this interpretation with the coordination statistics and distributions of angles between the neighbours of a given molecule, shown in Fig. \ref{coord}. The (mean) coordination number is given theoretically by integrating $4 \pi r^2 g(r)$ over the first peak, or practically by counting neighbours within a cutoff distance equal to the first minimum of the PDF (method as described in \cite{zircon}).These two methods give the same results (the same cutoff was used for calculating the angular distributions). At 1 kbar and 2.5 kbar, the coordination numbers, $n_c$, are close to 4, as expected in the tetrahedral-like structure and notably {\it increase} with temperature. Such an increase is anomalous (in a sense that $n_c$ and density usually decrease with increasing temperature) and is characteristic of water where higher temperatures disperse the tetrahedral structure, enabling more water molecules to move closer to a given molecule. We further observe that $n_c$ at low pressure increases up to about $T_{\rm F}$, at which point the transformation to the close-packed state is complete, in line with our earlier interpretation that dynamical crossover at the FL promotes the disappearance of the tetrahedral-like structure and enables densification into the closely-packed arrangement. The increase of $n_c$ up to $T_{\rm F}$ is followed by its decrease and the formation of {\it maxima} of $n_c$. The decrease of $n_c$ takes place in a closely-packed structure and is a generic effect of density decrease with temperature. These same qualitative results are found and thoroughly discussed at 1 kbar by Gorbaty \textit{et. al} in \cite{Gorbatyisobar}. Unlike at low pressure, no maxima are seen at higher pressure where the closely packed structure had already formed before the lowest temperature and where $n_c$ follows a generic decrease with temperature.

The transformation from low-density tetrahedral-like to a more closely-packed structure is also seen in the angular distribution in Fig. \ref{coord}c. The distribution has a peak at the tetrahedral angle of around $110^{\circ}$ at low temperature. As temperature increases, a new peak at around $60^{\circ}$ emerges and increases, representing close packing. The new peak reaches its maximum close to $T_{\rm F}$, corresponding to the largest number of closely-packed molecules. The angular distribution starts to flatten at yet higher temperatures, corresponding to the progressive loss of order in the structure. Representative structure snapshots with 4-fold and 6-fold coordinated water molecules are shown in Fig. \ref{cartoon}. We also observe the regions of high density (of about 15-30 \AA\ above the FL). This agrees with small-angle neutron scattering results in supercritical CO$_2$, which reported the appearance of droplets above the FL \cite{droplets}.

We now return to the PDFs at higher pressures. The high densities at these pressures mean that water can rearrange itself into a closer-packed arrangement with less dynamical assistance from the FL. For this reason the structural crossover at the FL is less pronounced at these pressures. The second PDF peaks disappear far before the FL at 5 kbar in Fig. \ref{pdfshigh}, suggesting that the transformation from tetrahedral to close-packing is nearly complete. At this pressure, the third peak begins its transition well before the FL as diffusive molecular motion becomes more prevalent. The FL marks the end of the transition of the third peak, where it begins to increase its radial position with temperature, typical of a simple liquid. At 10 kbar, the structure is close-packed below the FL (see the shape of the PDF in Fig. \ref{pdfshigh} and the coordination number in Fig. \ref{coord}), meaning the above crossover takes place within the close packed liquid.

\section{Conclusions}

These observations demonstrate the breadth of this structural crossover. At low pressure, the FL not only defines the region where water's tetrahedral structure becomes less pronounced than the close-packed structure, but also separates two regions of structural evolution: below the FL the evolution of structure is defined by the loss of tetrahedral order to close-packed order, above the FL the evolution is a generic loss of order due to increasing temperature and decreasing temperature. These results strongly imply that the loss of molecular oscillation and FL facilitate the known structural transformation in water. At higher pressures, the crossover can be seen in more subtle structural quantities - the PDF peak positions and heights, where again the FL separates two regions of distinct structural evolution.

These results importantly add to the previous experimental work revealing the structural crossover in liquid Ne \cite{ne}, CH$_4$ \cite{methane}, N$_2$ \cite{n2}, and CO$_2$ \cite{co2} at the FL and are important for two further reasons. First, our results serve as a stimulus and guide for future high pressure and temperature experiments aimed at elucidating supercritical water's phase diagram. Second, experimental data suggest that dissolving and extracting properties of supercritical fluids are optimised at the FL \cite{pre1}. Supercritical water is increasingly used in dissolving and environmental applications \cite{deben}, hence our results are industrially relevant.

\end{document}